\documentclass[a4paper,12pt]{article}
\usepackage{epsfig}
\usepackage[dvips,usenames]{color}
\usepackage{graphicx}
\usepackage{color}
\usepackage{colortbl}

\newlength{\dinwidth}
\newlength{\dinmargin}
\newcommand{\spur}[1]{\not\! #1 \,}

\begin{document}
\title{Probe the R-parity Violating
Supersymmetry Effects\\ in the $B^0_s-\bar{B}^0_s$ Mixing}
\author{Ru-Min Wang$^{1,2}$\thanks{E-mail address:
ruminwang@henannu.edu.cn }, G. R. Lu$^{1,2}$, En-Ke Wang$^{1}$  and
Ya-Dong Yang$^{2}$\thanks{E-mail address: yangyd@henannu.edu.cn }
 \\
 {\footnotesize {$^1$ \it Institute of Particle Physics,
 Huazhong Normal University,  Wuhan, Hubei 430070, P.R.China }}
  \\
{\footnotesize {$^2$ \it Department of Physics, Henan Normal
University, XinXiang, Henan 453007, P.R.China
 }}}
\maketitle
\begin{abstract}
The recent measurements of the $B_s$ mass difference $\Delta M_s$ by
the CDF and D{\O} collaborations are roughly consistent with the
Standard Model predictions, therefore, these measurements will
afford an opportunity to constrain new physics scenarios beyond the
Standard Model. We consider the impact of the R-parity violating
supersymmetry in the $B^0_s-\bar{B}^0_s$ mixing, and use the latest
experimental results of $\Delta M_s$ to constrain the size of the
R-parity violating tree level couplings in the $B^0_s-\bar{B}^0_s$
mixing. Then, using the constrained R-parity violating parameter
space from $\Delta M_s$, we show the R-parity violating effects on
the $B_s$ width difference $\Delta\Gamma_{s}$.
\end{abstract}

\vspace{1.5cm} \noindent {\bf PACS Numbers:  12.60.Jv, 11.30.Er,
  12.15.Mm, 14.40.Nd}

\newpage
Recently CDF and D{\O} collaborations have measured  the  mass
difference in the $B^0_s-\bar{B}^0_s$ system \cite{CDF,DO} with the
results
\begin{eqnarray}
\mbox{CDF:}&&\Delta M_s=(17.31^{+0.33}_{-0.18}\pm 0.07)/\mbox{ps},\label{exp} \\
\mbox{D{\O}:}&& 17/\mbox{ps}<\Delta M_s<21/\mbox{ps}~~~~
\mbox{(90\%~C.L.)}.
\end{eqnarray}

 The measurement of CDF collaboration  turned out to be surprisingly below
the Standard Model (SM) predictions obtained from other constraints
\cite{CKMfit,UTfit}
\begin{eqnarray}
\Delta M_s^{SM}(\mbox{UTfit})=(21.5\pm2.6)/\mbox{ps},~~~\Delta
M_s^{SM}(\mbox{CKMfit})=(21.7^{+5.9}_{-4.2})/\mbox{ps}.
\end{eqnarray}
A consistent though slightly smaller value is found for the mass
difference directly from its SM expression in later Eq.(\ref{SMDMs})
\begin{eqnarray}
\Delta
M_s^{SM}(\mbox{Direct})=(20.8\pm6.4)/\mbox{ps}.\label{SMderict}
\end{eqnarray}
with the input parameters collected in Table I. It's noted that this
prediction is sensitive to the value chosen for the non-perturbative
quantity $F_{B_s}\sqrt{B_{B_s}}$ and the CKM matrix element
$V_{ts}$, in this paper, we use their  values from
Refs.\cite{CKMfit,decayconstant}. The implication of $\Delta M_s$
measurements have already been studied in model independent approach
\cite{earlystudy}, MSSM models \cite{MSSMstudy}, $Z'$-model
\cite{Zpstudy}, Grand Unified Models \cite{GUT}.

 The SM
prediction in Eq.(\ref{SMderict}) suffers large uncertainties from
the hadronic parameters, nevertheless, the experimental data agree
fairly well with the SM value. Therefore, we can use the CDF
measurement to constrain new physics which may induce the b-s
transition. Effects of the R-parity violating (RPV) supersymmetry
(SUSY) on the neutral meson mixing have been discussed extensively
in the papers \cite{mixRPV,onlyschannel}. In this paper we will
consider the RPV SUSY effects at the tree level in the
$B^0_s-\bar{B}^0_s$ mixing by the latest experimental data. Using
the latest experimental data of $\Delta M_s$ and the theoretical
parameters, we obtain the new bound on the relevant RPV coupling
product. If there are RPV contributions to $\Delta M_s$, the same
new physics will also contribute to the width difference
$\Delta\Gamma_s$, and therefore we will use the constrained
parameter region to examine the RPV effects on $\Delta\Gamma_s$.

We first consider the SM contribution to the $B^0_s-\bar{B}^0_s$
mixing. The SM effective Hamiltonian for the $B^0_s-\bar{B}^0_s$
mixing is usually described by \cite{SMHeff}
\begin{eqnarray}
\mathcal{H}_{eff}^{SM}=\frac{G_F^2}{16
\pi^2}m_W^2|V_{ts}^*V_{tb}|^2\eta_{2B}S_0(x_t)[\alpha_s(\mu_b)]^{-6/23}
\left[1+\frac{\alpha_s(\mu_b)}{4\pi}J_5\right]\mathcal{O}+h.c.,\label{HSM}
\end{eqnarray}
with
\begin{eqnarray}
\mathcal{O}=(\bar{s}b)_{V-A} (\bar{s}b)_{V-A},\label{O}
\end{eqnarray}
 where $x_t=m^2_t/m^2_W$ and $\eta_{2B}$ is the QCD correction.

In terms of Eq.(\ref{HSM}), the mixing amplitude $M_{12}^{s}$ in the
SM, dominated by the top quark loop, is
\begin{eqnarray}
M_{12}^{s, SM}=\frac{\langle
B^0_s|\mathcal{H}_{eff}^{SM}|\bar{B}^0_s\rangle}{2m_{B_s}}.
\end{eqnarray}
Defining the renormalization group invariant parameter $B_{B_s}$ by
\begin{eqnarray}
&&B_{B_s}=B_{B_s}(\mu)[\alpha_s(\mu)]^{-6/23}
\left[1+\frac{\alpha_s(\mu)}{4\pi}J_5\right],\\
&&\langle B^0_s|\mathcal{O}|\bar{B}^0_s\rangle
\equiv\frac{8}{3}B_{B_s}(\mu)F^2_{B_s}m_{B_s}^2,
\end{eqnarray}
then, we have the $B_s$ mass difference in the SM
\begin{eqnarray}
\Delta M_s^{SM}&=&2\left|M_{12}^{s, SM}\right|\nonumber\\
&=&\frac{G_F^2}{6
\pi^2}m_W^2m_{B_s}|V_{ts}^*V_{tb}|^2\eta_{2B}S_0(x_t)\left(F_{B_s}\sqrt{B_{B_s}}\right)^2.
\label{SMDMs}
\end{eqnarray}

In the SM, the off-diagonal element of the decay
 width matrix $\Gamma^{s,SM}_{12}$  may be written as \cite{Benekembs}
\begin{eqnarray}
\Gamma^{s,SM}_{12}=-\frac{G^2_Fm^2_b}{24\pi
M_{B_s}}\left|V_{cb}V_{cs}^*\right|^2\left[G(x_c)\langle
B^0_s|\mathcal{O}|\bar{B}^0_s\rangle+G_2(x_c)\langle
B^0_s|\mathcal{O}_2|\bar{B}^0_s\rangle+\sqrt{1-4x_c}\hat{\delta}_{1/m}\right],
\end{eqnarray}
here $x_c=m_c^2/m_b^2$, $G(x_c)=0.030$ and $G_2(x_c)=-0.937$ at the
$m_b$ scale \cite{Benekembs}, and the $1/m_b$ corrections
$\hat{\delta}_{1/m}$ are given in \cite{1/mbcorrections}. The
operator $\mathcal{O}$ can be found in Eq.(\ref{O}), one now
encounters a second operator operator, $\mathcal{O}_2$, and thereby
another B-parameter $B^{(s)}_2(\mu)$
\begin{eqnarray}
\mathcal{O}_2=(\bar{s}b)_{S-P}(\bar{s}b)_{S-P},~~~\langle
B^0_s|\mathcal{O}_2(\mu)|\bar{B}^0_s\rangle=-\frac{5}{3}
\left(\frac{m_{B_s}}{\overline{m}_b(\mu)+\overline{m}_s(\mu)}\right)^2m^2_{B_s}f^2_{B_s}B^{(s)}_2(\mu).
\end{eqnarray}
The width difference between $B_s$ mass eigenstates is given by
\begin{eqnarray}
\Delta\Gamma_s^{SM}&=&2\left|\Gamma^{s,SM}_{12}\right| \nonumber\\
&=&\frac{G^2_Fm^2_b}{12\pi M_{B_s}}\left|V_{cb}V_{cs}^*\right|^2
\left[\frac{8}{3}G(x_c)B_{B_s}(\mu)F^2_{B_s}m_{B_s}^2 \right.\nonumber\\
&&\left.-\frac{5}{3}G_2(x_c)\left(\frac{m_{B_s}}{\overline{m}_b(\mu)+\overline{m}_s(\mu)}
\right)^2m^2_{B_s}f^2_{B_s}B^{(s)}_2(\mu)+\sqrt{1-4x_c}\hat{\delta}_{1/m}\right],
\end{eqnarray}
and the SM predicts $\Delta\Gamma_s^{SM}$ with the input parameters
in Table I
\begin{eqnarray}
\Delta \Gamma_s^{SM}(\mbox{Direct})=(0.07\pm0.03)/\mbox{ps}.
\end{eqnarray}
It's noted that the width difference  have been reviewed recently in
\cite{DGreview}.

 Now we turn to the RPV SUSY contributions to
the $B^0_s-\bar{B}^0_s$ mixing. In the most general superpotential
of the minimal supersymmetric Standard Model, the RPV superpotential
is given by \cite{RPVSW}
\begin{eqnarray}
\mathcal{W}_{\spur{R_p}}&=&\mu_i\hat{L}_i\hat{H}_u+\frac{1}{2}
\lambda_{[ij]k}\hat{L}_i\hat{L}_j\hat{E}^c_k+
\lambda'_{ijk}\hat{L}_i\hat{Q}_j\hat{D}^c_k+\frac{1}{2}
\lambda''_{i[jk]}\hat{U}^c_i\hat{D}^c_j\hat{D}^c_k, \label{rpv}
\end{eqnarray}
where $\hat{L}$ and $\hat{Q}$ are the SU(2)-doublet lepton and quark
superfields, $\hat{E}^c$, $\hat{U}^c$ and $\hat{D}^c$ are the
singlet superfields, while $i$, $j$ and $k$ are generation indices
and $c$ denotes a charge conjugate field.
\begin{figure}[ht]
\begin{center}
\includegraphics[scale=0.8]{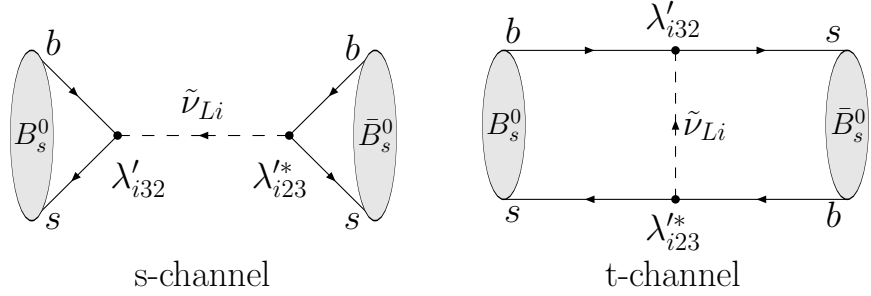}
\end{center}
\vspace{-0.9cm}
 \caption{ The RPV tree level contributions to the $B^0_s-\bar{B}^0_s$ mixing.}
 \label{RPVtreelevel}
\end{figure}

The $\lambda'$ couplings of Eq.(\ref{rpv}) make the
$B^0_s-\bar{B}^0_s$ mixing possible at the tree level through the
exchange of a sneutrino $\tilde{\nu}_i$ both in the s- and
t-channels displayed in Fig.\ref{RPVtreelevel}. The RPV tree level
contributions to $B^0_s-\bar{B}^0_s$ mixing are described by
\begin{eqnarray}
\mathcal{H}^{\spur{R_p}}_{eff}=\frac{1}{4}\sum_i
\frac{\lambda'_{i32}\lambda'^*_{i23}}{m^2_{\tilde{\nu}_{Li}}}
(\bar{s}b)_{S-P}(\bar{s}b)_{S+P}+h.c. , \label{EqHp}
\end{eqnarray}
where we have a new physics operator
\begin{eqnarray}
\mathcal{O}_4=(\bar{s}b)_{S-P}(\bar{s}b)_{S+P}, \label{NPoperators}
\end{eqnarray}
and we define the B-parameter as
\begin{eqnarray}
\langle B^0_s|\hat{\mathcal{O}}_4(\mu)|\bar{B}^0_s\rangle&=&2
\left(\frac{m_{B_s}}{\overline{m}_b(\mu)+\overline{m}_s(\mu)}\right)^2m_{B_s}^2F_{B_s}^2B^{(s)}_4(\mu).
\end{eqnarray}
Note that the expectation values are scaled by factor of $2m_B$ over
those given in some literature due to our different normalization of
the meson wave functions. It is trivial to check that both
conventions yield the same values for physical observables.

The RPV mixing amplitude $M_{12}^{s,\spur{R_p}}$ is
\begin{eqnarray}
M_{12}^{s, \spur{R_p}}&=&\frac{\langle
B^0_s|\mathcal{H}_{eff}^{\spur{R_p}}|\bar{B}^0_s\rangle}{2m_{B_s}}\nonumber\\
&=&\sum_i
\frac{\lambda'_{i32}\lambda'^*_{i23}}{m^2_{\tilde{\nu}_{Li}}}\frac{1}{4}
\left(\frac{m_{B_s}}{m_b(\mu)+m_s(\mu)}\right)^2m_{B_s}F_{B_s}^2B^{(s)}_4(\mu),
\end{eqnarray}

Given the expressions above, we now write the total $B_s$ mass
difference included both SM
 and  RPV contributions
\begin{eqnarray}
\Delta M_s=2|M^s_{12}|,
\end{eqnarray}
with
\begin{eqnarray}
M^s_{12}&=&M^{s,SM}_{12}+M_{12}^{s,\spur{R_p}}\nonumber\\
&=&M^{s,SM}_{12}(1+ze^{i\theta}),
\end{eqnarray}
where the parameters $z$ and $\theta$ give the relative magnitude
and relative phase of the RPV contribution, i.e. $z\equiv
\left|M_{12}^{s\spur{R_p}}/M^{s,SM}_{12}\right|$ and
$\theta\equiv$arg$\left(M_{12}^{s,\spur{R_p}}/M^{s,SM}_{12}\right)$.

The $B_s$ width difference beyond the SM has been studied in
Refs.\cite{XZZ,Grossman}. If there are RPV contributions to $\Delta
M_s$, the same new physics will also contribute to the $B_s$ width
difference.
 The width difference including the RPV contributions is given by
\cite{Grossman}
\begin{eqnarray}
\Delta\Gamma_s=\frac{4|Re(M^s_{12}\Gamma^{s*}_{12})|}{\Delta M_s}
=2|\Gamma^s_{12}|\cdot|\mbox{cos}\phi_m|,
\end{eqnarray}
where $\phi_m=$arg$(1+ze^{i\theta})$, and $\phi_m=0$ turns out to be
an excellent approximation in the SM. The effect of NP on the
off-diagonal element of the decay
 width matrix $\Gamma^s_{12}$ is anticipated to be negligibly small,
hence $\Gamma^s_{12}=\Gamma^{s,SM}_{12}$ holds as a good
approximation \cite{Gamma12}.

 We now perform numerical
calculation and show the constraint imposed by the measurement of
$\Delta M_s$ only or both $\Delta M_s$ and $\Delta \Gamma_s$. The
values of the input parameters used in this paper are collected in
Table I, and we will use the input parameters and the experimental
data which vary randomly within $1\sigma$ variance.

\begin{table}[ht]
\centerline{\parbox{12.3cm}{Table I: Values of the theoretical
quantities as input parameters. }} \vspace{0.3cm}
\begin{center}
\begin{tabular}{lc}\hline\hline
$m_W=80.403\pm
0.029~GeV,~~~m_{B_s}=5.3696\pm 0.0024~GeV,$&\\
$\overline{m}_b(\overline{m}_b)=4.20\pm0.07~GeV,~~~\overline{m}_s(2GeV)=0.095\pm0.025~GeV,$&\\
$m_t=174.2\pm3.3~GeV,~~~m_b=4.8~GeV$.&\cite{PDG2006}\\\hline
$A=0.818^{+0.007}_{-0.017},~~~\lambda=0.2272\pm0.0010.$&\cite{PDG2006}\\\hline
$\eta_{2B}=0.55\pm0.01.$&\cite{eta2B}\\\hline
$F_{B_s}\sqrt{B_{B_s}}=0.262\pm
0.035~GeV,~~~F_{B_s}=0.230\pm0.030~GeV.$&\cite{decayconstant}\\\hline
$B_{2}^{(s)}(m_b)=0.832\pm0.004,~~~B_{4}^{(s)}(m_b)=1.172^{+0.005}_{-0.007}.$
&\cite{Bparameter}\\\hline\hline
\end{tabular}
\end{center}
\end{table}
\begin{figure}[ht]
\begin{center}
\includegraphics[scale=0.45]{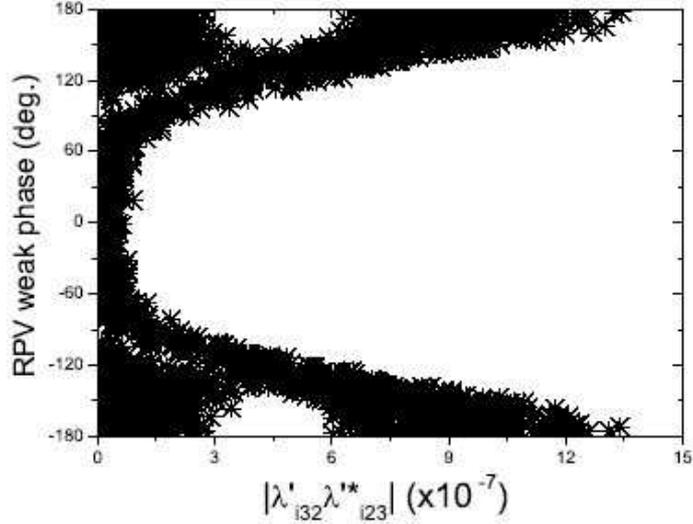}
\end{center}
\vspace{-0.9cm}
 \caption{\small Allowed parameter space for
 $\lambda'_{i32}\lambda'^*_{i23}$ constrained by  the experimental data of $\Delta M_s$.}
 \label{bounds1}
\end{figure}

We calculate the contributions of Eq.(\ref{EqHp}) to $\Delta M_s$
and require it not to exceed the corresponding experimental data in
Eq.(\ref{exp}). The random variation of the parameters subjecting to
the constraint leads to the scatter plot shown in Fig.\ref{bounds1}.

We can see that there are three possible bands of solutions in
Fig.\ref{bounds1}. The two bands are for the modulus of RPV weak
phase $(\phi_{\spur{R_p}})\in[\frac{5}{9}\pi,\pi]$ and
$|\lambda'_{i32}\lambda'^*_{i23}|\leq3.2\times10^{-7}$. The other
band is for $\phi_{\spur{R_p}}\in[-\pi,\pi]$ and
$|\lambda'_{i32}\lambda'^*_{i23}|\leq1.4\times10^{-6}$,
$|\phi_{\spur{R_p}}|$ is increasing with
$|\lambda'_{i32}\lambda'^*_{i23}|$ in this band. We get a very
strong bound on the magnitudes of the RPV coupling product
$\lambda'_{i32}\lambda'^*_{i23}$ from $\Delta M_s$
\begin{eqnarray}
|\lambda'_{i32}\lambda'^*_{i23}|\leq 1.4\times 10^{-6}\times
\left(\frac{100~GeV}{m_{\tilde{\nu}_i}}\right)^2.
\end{eqnarray}
For comparison, we will use the existing bounds on these single
coupling in Refs.\cite{RPVbound,EDMbound,Allanach} to compose the
corresponding bounds on the quadric coupling products with the
superpartner mass of 100 $GeV$. In the RPV SUSY model, the strongest
bound for this coupling is $|\lambda'_{i32}\lambda'^*_{i23}|\leq
1.4\times 10^{-3}$ in Ref.\cite{RPVbound}, and some bounds are
obtained $|\lambda'_{132}\lambda'^*_{123}|\leq 1.0\times 10^{-11}$
and $|\lambda'_{232}\lambda'^*_{223}|\leq 1.0\times 10^{-3}$ by the
experimental upper limits on the electric dipole moment's of the
fermions in Ref.\cite{EDMbound}. In addition, in the RPV mSUGRA
model, Allanach $et~ al.$ have obtained quite strong upper bound:
$|\lambda'_{i32}\lambda'^*_{i23}|\leq 2.6\times 10^{-9}$ at the
$M_{GUT}$ scale and $|\lambda'_{i32}\lambda'^*_{i23}|\leq 2.2\times
10^{-8}$ at the $M_{Z}$ scale \cite{Allanach}, so their constraints
from neutrino masses are stronger than ours from the
$B^0_s-\bar{B}^0_s$ mixing. However, we note that the constraints on
$\lambda'$ from neutrino masses would depend on the explicit
neutrino mass models with trilinear couplings only, bilinear
couplings only, or both \cite{RPVbound}.
\begin{figure}[ht]
\begin{center}
\begin{tabular}{cc}
\includegraphics[scale=0.4]{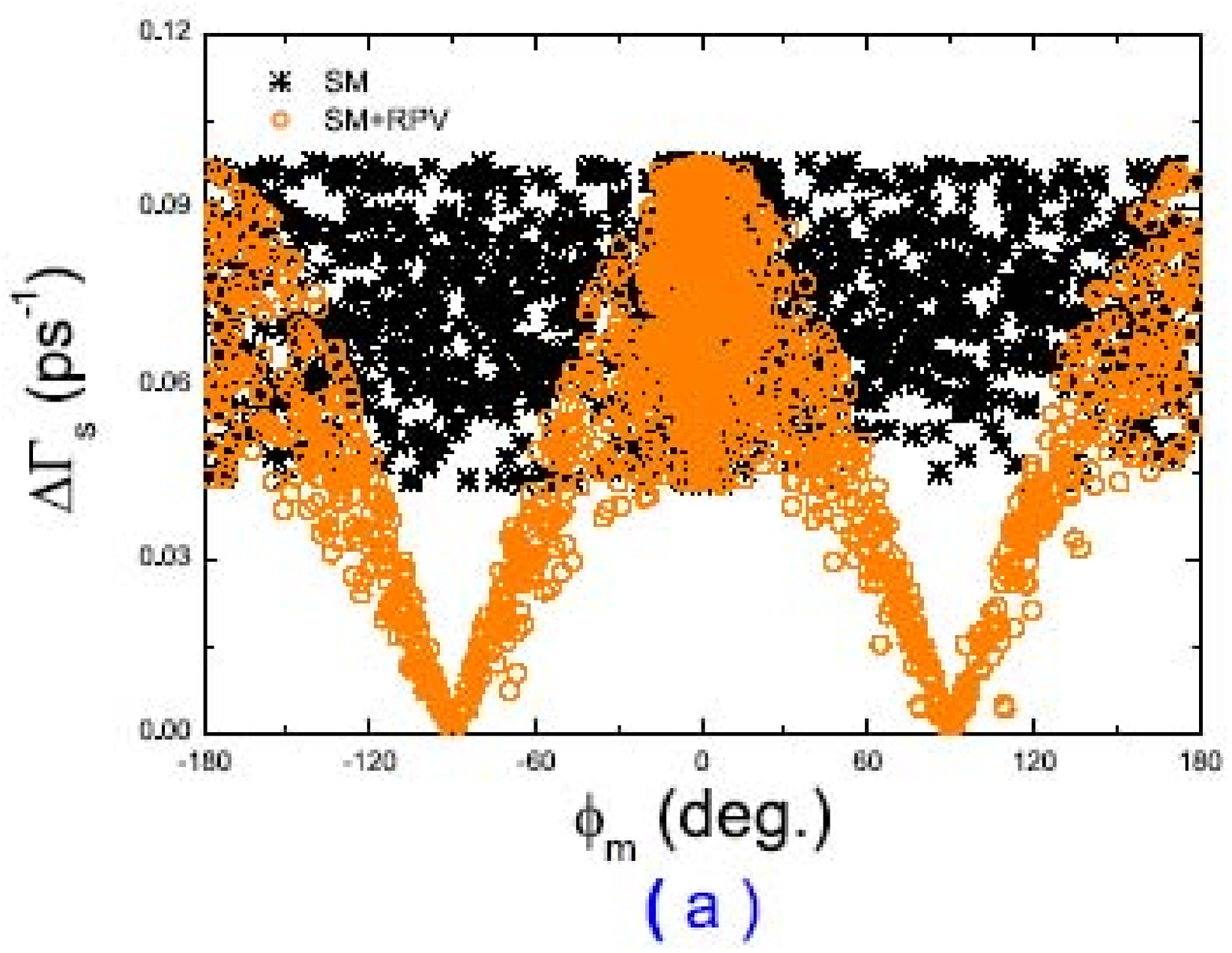}&
\includegraphics[scale=0.38]{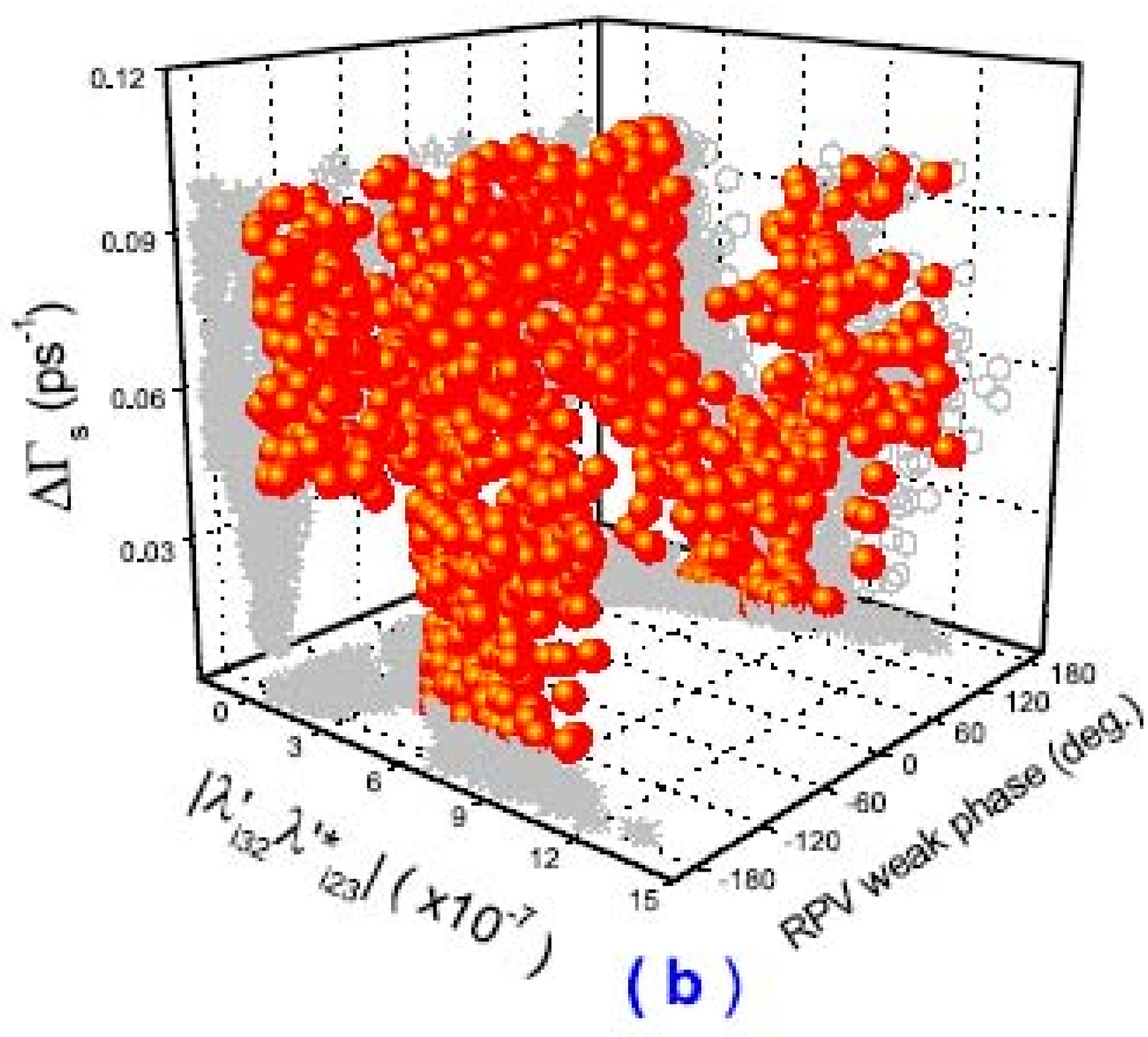}
\end{tabular}
\end{center}
\vspace{-0.9cm}
 \caption{ The RPV tree level contributions to the $\Delta\Gamma_s$.}
 \label{DG}
\end{figure}

Using the constrained  parameter space  from $\Delta M_s$ as shown
in Fig.\ref{bounds1}, one can predict the RPV effects on the $B_s$
width difference $\Delta\Gamma_s$. Our
 predictions of $\Delta\Gamma_s$ are displayed in Fig.\ref{DG}. From
Fig.\ref{DG}(a), we find that $\phi_m$ can have any value
  from $-\pi$ to $\pi$, as discussed in Ref.\cite{Grossman}, the RPV
contributions to the mixing could reduce $\Delta \Gamma_s$ relative
to the SM prediction, and $\Delta\Gamma_s$ lies between
$0.00/\mbox{ps}$ and $0.10/\mbox{ps}$.
 We present correlation between $\Delta\Gamma_s$ and the parameter
space of $\lambda'_{i32}\lambda'^*_{i23}$ by the three-dimensional
scatter plot in Fig.\ref{DG}(b).
 We also give projections on three vertical
 planes, where the $|\lambda'_{i32}\lambda'^*_{i23}|$-$\phi_{\spur{R_p}}$ plane displays the
 constrained
 region of $\lambda'_{i32}\lambda'^*_{i23}$ as the plot of Fig.\ref{bounds1}.
 It's shown that $\Delta\Gamma_s$ is  decreasing first and then
 increasing
 with $|\lambda'_{i32}\lambda'^*_{i23}|$
 on the $\Delta\Gamma_s$-$|\lambda'_{i32}\lambda'^*_{i23}|$ plane.
 From the $\Delta\Gamma_s$-$\phi_{\spur{R_p}}$ plane,
 we can see that $\Delta\Gamma_s$ may be reduced to zero when
 $|\phi_{\spur{R_p}}|$ lies in $[\frac{2}{3}\pi,\frac{8}{9}\pi]$.
\begin{figure}[htbp]
\begin{center}
\includegraphics[scale=0.45]{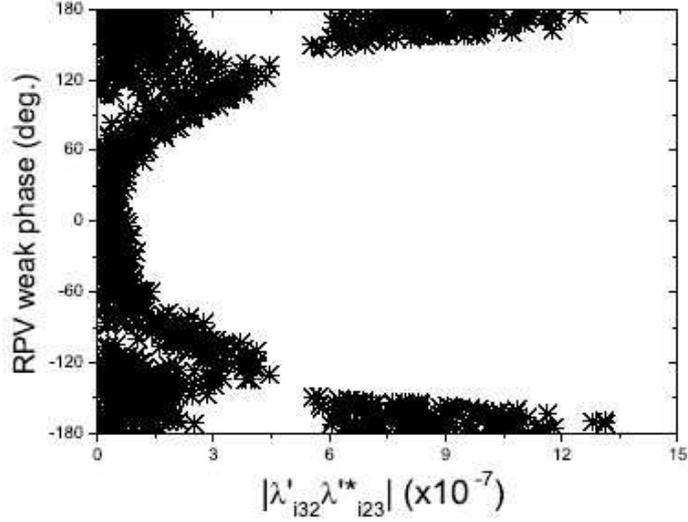}
\end{center}
\vspace{-0.9cm}
 \caption{\small Allowed parameter space for
 $\lambda'_{i32}\lambda'^*_{i23}$ constrained by  the data of $\Delta M_s$ and $\Delta \Gamma_s$.}
 \label{bounds2}
\end{figure}

The present experimental data of the $B_s$ width difference have a
large error, and we obtain the averaged value from
\cite{PDG2006,DGexp}
\begin{eqnarray}
\Delta\Gamma_s=(0.22\pm0.09)/\mbox{ps}. \label{DGdata}
\end{eqnarray}
Now we add the experimental constraint of $\Delta\Gamma_s$ to
 the allowed space of $\lambda'_{i32}\lambda'^*_{i23}$. We can not
 get the solution to the experimental data of $\Delta\Gamma_s$ at $1\sigma$ level. If
 $\Delta\Gamma_s$ is varied randomly within $2\sigma$ variance, we can obtain the
scatter plot as exhibited in Fig.\ref{bounds2}. Comparing
Fig.\ref{bounds2} with Fig.\ref{bounds1}, we can see that the
experimental bound on $\Delta\Gamma_s$ shown in Eq.(\ref{DGdata})
obviously excludes  the region $4.4\times
10^{-7}<|\lambda'_{i32}\lambda'^*_{i23}|<5.5\times 10^{-7}$. The
stronger limit on $|\lambda'_{i32}\lambda'^*_{i23}|$ from $\Delta
M_s$ and $\Delta \Gamma_s$ than the one from $\Delta M_s$ only is
 obtained
 \begin{eqnarray}
&&|\lambda'_{i32}\lambda'^*_{i23}| \leq 4.4\times 10^{-7}\times
\left(\frac{100~GeV}{m_{\tilde{\nu}_i}}\right)^2,\\
\mbox{and}&& \nonumber\\
&&|\lambda'_{i32}\lambda'^*_{i23}|\in[5.5,13.1]\times 10^{-7}\times
\left(\frac{100~GeV}{m_{\tilde{\nu}_i}}\right)^2.
\end{eqnarray}

 In summary,
we have studied the RPV tree level effects in the
$B^0_s-\bar{B}^0_s$ mixing with the current experimental
measurements. As shown, using the latest experimental data of
$\Delta M_s$ and the theoretical parameters, we have obtained the
allowed space of the RPV coupling product
$\lambda'_{i32}\lambda'^*_{i23}$, the upper bound on the magnitude
of $\lambda'_{i32}\lambda'^*_{i23}$ has been greatly improved over
the existing bounds obtained from the RPV SUSY. Then, we have
examined the RPV effects on $\Delta \Gamma_s$ by the constrained
region of $\lambda'_{i32}\lambda'^*_{i23}$ from $\Delta M_s$, and we
have found that the RPV contributions to the mixing could reduce
$\Delta \Gamma_s$ relative to the SM prediction.  Finally, using the
experimental data of $\Delta M_s$ and $\Delta \Gamma_s$,  we have
obtained stronger bound than the one from $\Delta M_s$ only on
$\lambda'_{i32}\lambda'^*_{i23}$. In addition, we stress that once
LHC is turned on, with the anticipated production of $10^{12}$
$b\bar{b}$ per year, the measurements of $\Delta M_s$ and $\Delta
\Gamma_s$ will be much more accurate, then the allowed parameter
space for $\lambda'_{i32}\lambda'^*_{i23}$ will be significantly
shrunken or ruled out. \vspace{-0.7cm}
\section*{Acknowledgments}
Ru-Min Wang wish to thank Alexander Lenz for useful suggestions. The
work is supported by National Natural Science Foundation of China
under contract Nos.10305003, 10475031, 10440420018 and by MOE of
China under projects NCET-04-0744, SRFDP-20040511005,
CFKSTIP-704035. 
\end{document}